\documentclass[aps,prl,superscriptaddress,showpacs,balancelastpage,10pt,reprint]{revtex4-1} 

\usepackage{graphicx}

\usepackage{amsmath}
\usepackage{amsfonts}
\usepackage{amssymb}

\begin{document}

\title{Phase space tweezers for tailoring cavity fields by quantum Zeno dynamics}

\author{J.M. Raimond}
\author{C. Sayrin}
\author{S. Gleyzes}
\author{I. Dotsenko}
\author{M. Brune}
\affiliation{Laboratoire Kastler Brossel, CNRS, ENS, UPMC-Paris 6, 24 rue Lhomond, 75231 Paris, France}
\author{S. Haroche}
\affiliation{Laboratoire Kastler Brossel, CNRS, ENS, UPMC-Paris 6, 24 rue Lhomond, 75231 Paris, France}
\affiliation{Coll\`{e}ge de France, 11 place Marcelin Berthelot,
75231 Paris, France}
\author{P. Facchi}
\affiliation{Dipartimento di Matematica and MECENAS, Universit\`a
di Bari,
        I-70125  Bari, Italy}
\affiliation{INFN, Sezione di Bari, I-70126 Bari, Italy}
\author{S. Pascazio} 
\affiliation{Dipartimento di Fisica and MECENAS, Universit\`a di Bari,
        I-70126  Bari, Italy}
\affiliation{INFN, Sezione di Bari, I-70126 Bari, Italy}

\date{\today{}}

\begin{abstract}
We discuss an implementation of Quantum Zeno Dynamics in a Cavity Quantum Electrodynamics experiment. By performing repeated unitary operations on atoms coupled to the field, we restrict the field evolution in chosen subspaces of the total Hilbert space. This procedure leads to promising methods for tailoring non-classical states. We propose to realize `tweezers' picking a coherent field at a point in phase space and moving it towards an arbitrary final position without affecting other non-overlapping coherent components. These effects could be observed with a state-of-the-art apparatus.
\end{abstract}


\maketitle

In the Quantum Zeno effect (QZE)~\cite{Zenoproposal}, repeated
projective measurements block the evolution of a system in a
non-degenerate eigenstate of the measured observable. It has been
observed on two-level systems~\cite{Zenotwolevel} and on an
harmonic oscillator in a Cavity Quantum Electrodynamics (CQED)
experiment~\cite{Bernu08}. A Quantum Zeno dynamics (QZD)~\cite{Facchi02} 
takes place when the system is not confined to a
single state, but rather evolves under the action of its free
hamiltonian $H$ in a multidimensional subspace of its Hilbert
space. This can be achieved either by repeated measurements of an
observable with degenerate eigenvalues, or by repeated actions of
a unitary kick $U_K$ with multi-dimensional invariant subspaces,
the two procedures being physically equivalent~\cite{Facchi04}. We
focus here on the latter case, related to the so-called
`bang-bang' control~\cite{BangBang} and NMR manipulation
techniques~\cite{NMR}.

The system evolution is stroboscopic, alternating small free
evolution steps described by $U(\delta t)=\exp(-iH\delta t/\hbar)$
with $U_K$ kicks. The succession of $N$ steps (fixed duration
$t=N\delta t$) corresponds to the unitary
$U_{Z}(N)=[U_KU(t/N)]^N$. It is, in the $N\rightarrow\infty$
limit, the evolution under an effective Hamiltonian
$H_{Z}=\sum_\mu P_\mu H P_\mu$ where the $P_\mu$'s are the
projectors on the invariant subspaces of $U_K$~\cite{Facchi02}. By
choosing properly  $U_K$ (or equivalently the repeatedly measured
observable), one can tailor the system evolution, leading to
decoherence control~\cite{Facchi04}, state purification~\cite{Nakazato04} 
and quantum gates implementation~\cite{Shao09}.
QZD can also inhibit entanglement between subsystems, making a
quantum evolution semi-classical~\cite{Rossi09}.

In this Letter, we propose a CQED implementation of the bang-bang
QZD control. We exploit the non-linearity of the atom-cavity
system~\cite{Exploring06} to implement a photon-number-selective
$U_K$. The field dynamics in its Hilbert space $\cal H$ is
confined in two disconnected subsets ${\cal H}_{<s}$ and ${\cal
H}_{>s}$, corresponding to photon numbers smaller or larger than a
preset value $s$. This leads to novel methods of non-classical
field states preparation and tailoring. We propose a `phase space
tweezer' picking selectively a coherent state component of a
quantum superposition and moving it at will in phase space
independently from the other components.

\begin{figure}
\includegraphics[width=6.5cm]{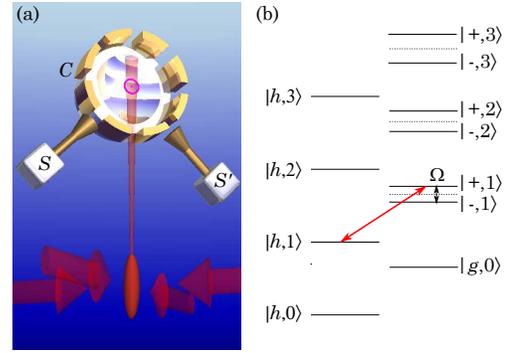}
  \caption{(a) Proposed experimental scheme. A slow atomic beam
extracted from a 2D-Magneto Optical Trap (bottom) forms an atomic
fountain with atoms nearly at rest in the center of the
high-quality microwave Fabry-Perot cavity $C$ (only one mirror
shown). Sources $S$ and $S'$ address respectively the dressed
atomic levels and the cavity mode. Electrodes around the cavity
mirrors generate the electric fields preparing the circular state
shown in the center. (b) Scheme of the dressed atomic levels. The
arrow indicates the photon-number selective transition addressed
by $S$ for $s=1$.}
  \label{scheme}
\end{figure}

Our proposal could be implemented in a microwave CQED experiment
with circular Rydberg atoms and a superconducting millimeter-wave
cavity~\cite{Exploring06}.  The cavity $C$ [Fig. \ref{scheme}(a)]
is crossed by a slow beam of Rubidium ground state atoms, in an
`atomic fountain' arrangement. Close to their turning point, atoms
are nearly at rest. One of them is promoted to the circular level
$h$ (principal quantum number 49) using static and r.f. electric
fields. This operation does not change the state of $C$, tuned on
resonance with the 51.1~GHz transition between $g$ and $e$ (50 and
51 circular states). The analysis that follows pertains to the
dynamics of this single atom and the cavity.

\begin{figure*}
\includegraphics[width=16cm]{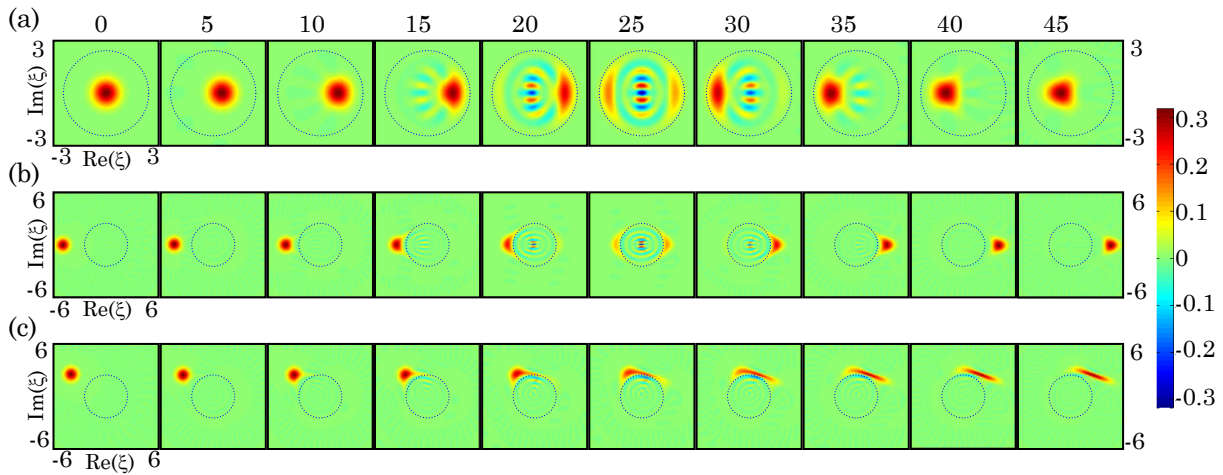}
\caption{(a) QZD dynamics in ${\cal H}_{<6}$. Ten snapshots of the
field Wigner function $W(\xi)$ obtained after a number of steps
indicated above each frame. The cavity is initially in its vacuum
state, $s=6$ and $\beta=0.1$. The EC is plotted as a blue dashed
line. (b) QZD dynamics in ${\cal H}_{>6}$. Same as (a) with an
initial $\alpha=-5$ amplitude. (c) Same as (b), with an initial
amplitude $\alpha=-4+i\sqrt{6}$. In (b) and (c) the successive
frames correspond to the same step numbers as in (a).}
  \label{confine}
\end{figure*}

The source $S$  drives the $h\rightarrow g$ transition near
54.3~GHz. It does not couple directly to the off-resonant cavity.
It realizes $U_K$ by probing the eigenstates of the atom/cavity
system [Fig. \ref{scheme}(b)]. The energies of the $|h,n\rangle$
states (atom in $h$ with $n$ photons) do not depend on the
atom-cavity coupling, since $C$ is far off-resonance from the
$h\rightarrow g$ transition. The source $S$ couples $|h,n\rangle$
to the dressed states
$|\pm,n\rangle=(|e,n-1\rangle\pm|g,n\rangle)/\sqrt{2}$, which are
superpositions of the degenerate uncoupled states $|e,n-1\rangle,\
|g,n\rangle$. The splitting between dressed states is
$\Omega\sqrt{n}$ where $\Omega$ is the vacuum Rabi frequency.
Hence, the $|h,n\rangle\rightarrow|+,n\rangle$ transition
frequency depends upon $n$. We tune $S$ to perform on the
$|h,s\rangle\rightarrow|+,s\rangle$ transition a $2\pi$ Rabi pulse
whose amplitude is weak enough (and its duration correspondingly
long enough) not to appreciably affect $|h,n\rangle$ with
$n\not=s$. It results in the transformation
$|h,n\rangle\rightarrow (-1)^{\delta_{ns}}|h,n\rangle$. The atom
always ends up in $h$ while the field experiences $U_K=U_s$ with
$U_s=\openone-2|s\rangle\langle s|$. Such a photon-number dependent Rabi
pulse~\cite{Franca87} was used with $s=1$ for a single-photon QND
detection~\cite{QND99} and for a CNOT gate in CQED~\cite{Gate99}.

The free cavity dynamics is produced by the source $S'$,
resonantly coupled with $C$ and acting during time intervals
$\delta t$ between two $U_s$ operations. Being not resonant with
the atom in $h$, $S'$ leaves it unaffected. The free evolution is
described by the Hamiltonian  $H=-i({\cal E}^*a-{\cal
E}a^\dagger)$, where $\cal E$ is the source amplitude and $a$
($a^\dagger$) the photon annihilation (creation) operator. We use
an interaction representation eliminating the field phase rotation
at cavity frequency. The unitary $U(\delta t)$ is the displacement
$D(\beta)=\exp(\beta a^\dagger-\beta^* a)$, with $\beta={\cal
E}\delta t/\hbar\ll 1$. After $p$ repetitions of $U_sU(\delta t)$,
the cavity state can be reconstructed~\cite{Wigner08}.

The eigenvalues of $U_s$ are $-1$ and $+1$. The former corresponds
to the one-dimension eigenspace ${\cal H}_s$, generated by
$|s\rangle$ (projector $P_{s}$). The latter is associated to the
direct sum of ${\cal H}_{<s}$ (projector $P_{<s}$), generated by
the Fock states $|0\rangle,\ldots,|s-1\rangle$, and ${\cal
H}_{>s}$ (projector $P_{>s}$) generated by Fock states above
$|s\rangle$. The projectors $P_\mu$ ($\mu=+,-$) are thus
$P_-=P_{s}$ and $P_+=P_{<s}+P_{>s}$.

Since $H$ is a linear combination of $a$ and $a^\dagger$, $H_{Z}$
reduces to $P_{<s} H P_{<s}+ P_{>s} H P_{>s}$. Under the QZD,
field states restricted to ${\cal H}_{<s}$ and  ${\cal H}_{>s}$
remain confined in these subspaces, $|s\rangle$ realizing a hard
`wall' between them. This wall induces remarkable features in the
evolution. If we start from the vacuum in $C$ with $s=1$, the
system remains inside ${\cal H}_{<1}$ i.e. in $|0\rangle$. We
recover the QZE~\cite{Bernu08}.

We have simulated this QZD procedure~\cite{Quopackage}. Fig.
\ref{confine}(a) presents 10 snapshots of the field Wigner
function, $W(\xi)$, separated by intervals of 5 steps, for $s=6$
and $\beta=0.1$. The field starts from $|0\rangle \in {\cal
H}_{<6}$. Its amplitude first increases along the real axis (free
dynamics). Between 15 and 20 steps, the amplitude reaches $\simeq
2$ and QZD comes into play. The coherent state `collides' on the
$U_s$-induced `wall', materialized in phase space as an `Exclusion
Circle' (EC) of radius $\sqrt{6}$ (dashed line in Fig.
\ref{confine}). The field amplitude stops growing and undergoes a
progressive $\pi$ phase shift between steps 20 and 30. At step 25,
the field is in  a `cat state', quantum superposition of two
components with opposite phases. The fringing feature inside the
EC is the signature of the quantum coherence of this
superposition. This cat contains only odd photon numbers
(probabilities for 5, 3 and 1 photons are $0.63$, $0.31$ and
$0.03$). At step 35, the field state is nearly coherent with an
amplitude close to -2. It then resumes its motion from left to
right along the real axis, going through $|0\rangle$ again (around
step 45) and heading towards its next `collision' with the EC.

\begin{figure}
\includegraphics[width=7cm]{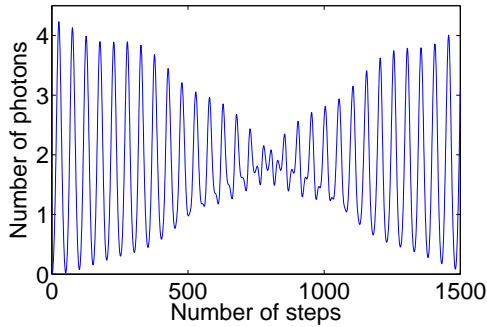}
\vspace{0cm} \caption{Energy of the field as a function of the
number of steps for a QZD dynamics inside ${\cal H}_{<6}$.
Conditions are the same as for Fig. \ref{confine}(a), which
presents snapshots of the first oscillation.}
  \label{Oscillations}
\end{figure}

Fig. \ref{Oscillations} presents the long-term evolution of the
field energy. During the first few hundred steps, the oscillations
reveal the quasi-periodic motion of the field inside the EC. State
distortions, however, accumulate and damp the oscillations, whose
contrast nearly vanishes after 800 steps. Since there is only a
finite set of frequencies in $P_{<6} H P_{<6}$, we
observe at longer times a quantum revival~\cite{Exploring06}. The
energy oscillations resume and the field comes back to an
oscillating coherent state periodically colliding with the EC as
described above.

Fig. \ref{confine}(b) illustrates QZD in ${\cal H}_{>s}$ with
snapshots of the field Wigner function for $s=6$ and an initial
coherent state $|\alpha=-5\rangle$. The field collides on the EC
after 20 steps. It undergoes a QZD-induced $\pi$ phase shift
being, after 25 steps, in a cat state. After 30 steps, the state
is again nearly coherent with a positive amplitude. It resumes its
motion along the real axis. After 45 steps, its amplitude is
slightly larger than 4.5. It would be $-0.5$ in the case of free
dynamics. The QZD-induced phase inversion accelerates the
`propagation' in phase space. This opens interesting possibilities
when the initial amplitude is such that the field state collides
tangentially on the EC. The parts of the Wigner function that come
closest to the EC propagate faster than others. The state is
distorted and ends up strongly squeezed [Fig. \ref{confine}(c)].

QZD can be generalized to ECs centered at an arbitrary
point $\gamma$ in phase space by changing the kick operator $U_K$
from $U_s$ to $U_s(\gamma)=D(\gamma)U_s D(-\gamma)$ (these
displacements $\gamma$ being also performed by $S'$). After $p$ steps, the
global evolution operator is
$U_{Z}(s,\gamma,p)=[U_s(\gamma)D(\beta)]^p$ which can be
expressed, using displacement operator commutation relations, as
$U_{Z}(s,\gamma,p)=D(\gamma)U_{Z}(s,0,p)D(-\gamma)\exp[2ip\Im(\beta\gamma^*)]$.
Up to a topological phase, the state after $p$ steps is
equivalently obtained by first displacing the field by $-\gamma$,
performing $p$ QZD steps in an EC centered at origin and finally
displacing back the field by $\gamma$.

This leads to the concept of phase space tweezer. Applying this
procedure with $s=1$ to an initial cat state
$|\gamma\rangle+|\alpha\rangle$
($\langle\gamma|\alpha\rangle\approx 0$), we can selectively block
the evolution of $|\gamma\rangle$ while leaving the other
component free to evolve. After $N$ steps, we get the `stretched'
cat  $|\gamma\rangle+D(N\beta)|\alpha\rangle=|\gamma\rangle
+\exp[iN\Im(\beta\alpha^*)]|\alpha+N\beta\rangle$.

In an interesting variant, the position $\gamma_p$ of the tweezer
at step $p$ is changed by a small amount at each step
($|\gamma_{p+1}-\gamma_p|\ll 1$), while $\beta$ is set to zero (no
free evolution). The sequence of $\{\gamma_p\}$ defines a trajectory
$\cal T$ in phase space followed by the center of the EC. A
coherent state with the initial amplitude $\alpha_i=\gamma_1$
follows adiabatically this trajectory and becomes, after the
$p$-th step, a coherent state with amplitude $\alpha_p=\gamma_p$.
We realize in this way a tweezer which moves one selected coherent
state, while not affecting the evolution of all the coherent
states whose amplitude remains  away from $\cal T$.

A combination of tweezers can move all the components of a
superposition of non-overlapping coherent states from arbitrary
initial to final positions. An obvious method is to grasp them one
by one, driving them from their initial to their final position
(taking care to move them so that different components never
overlap). The tweezers can also be used in parallel by applying
incremental motions alternatively on each component.

\begin{figure}
\includegraphics[width=7.5cm]{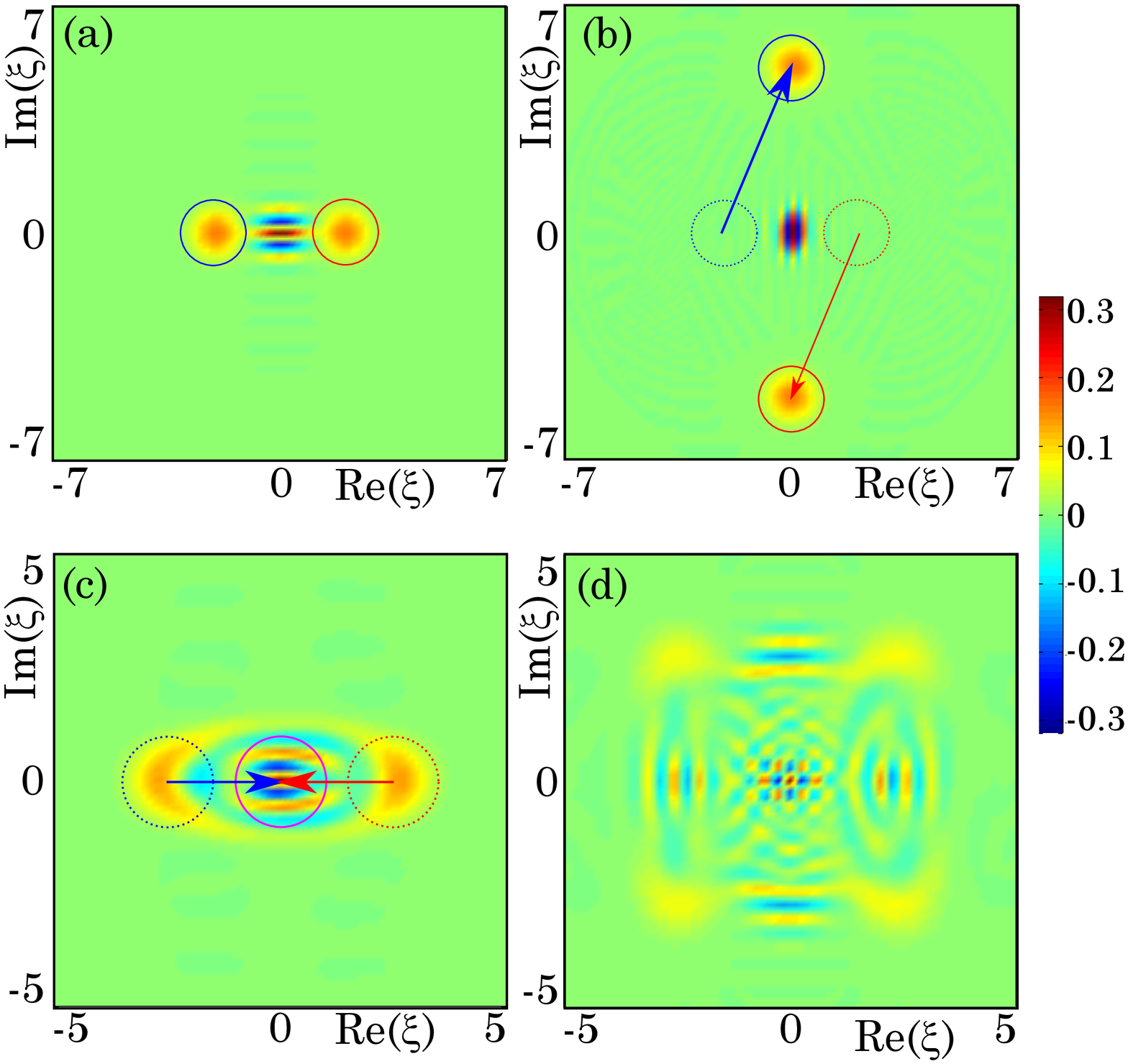}
\caption{(a)-(b) Initial and final Wigner functions $W(\xi)$ for a
phase space tweezer operation. The first step ECs are depicted as
solid lines in (a) and dotted lines in (b), the final ECs as solid
lines in (b). The arrows in (b) indicate the two EC centers
trajectories.  (c) Cat created by a vacuum state crush operation.
Initial (dotted lines) and final (solid line) ECs are plotted,
with arrows indicating the motion of their centers. (d)
Four-component cat created by three successive crushing
operations.}
 \label{tweezer}
\end{figure}

Fig. \ref{tweezer} illustrates this procedure for a two-component
cat, initially $|\alpha\rangle+|-\alpha\rangle$ with $\alpha=2$.
It is turned in 100 steps (50 on each component) into
$|\alpha'\rangle+|-\alpha'\rangle$ with $\alpha'=5i$. Panels (a)
and (b) present the initial and final Wigner functions. The final
fidelity is $98.8$\% with respect to the expected cat. It remains
greater than 68\% if the operation is performed in 20 steps only,
exhibiting the robustness of this adiabatic procedure, which is
promising for experimental implementations.

The initial cat state can be generated in various ways using
dispersive~\cite{Decoherence96} or resonant~\cite{Resonantcat}
atoms. An adapted tweezer procedure also prepares from $|0\rangle$
mesoscopic superpositions which are approximations of these cat
states. We `crush' the vacuum between two $s=1$ ECs whose centers,
initially at $\pm 2.5$, move simultaneously towards each other in
200 steps, until they reach the origin 
(the wavefunction remaining outside both ECs). Fig. \ref{tweezer}(c)
presents the final Wigner function. The state is a superposition
of two well-separated components (average energy 6.4 photons). Its
fidelity with respect to a $|\alpha\rangle+|-\alpha\rangle$ cat
with the same average energy is 42\%. The final components can be
crushed again and so on, leading to a superposition state with an
arbitrary number of components. Fig. \ref{tweezer}(c) presents the
Wigner function of a four-component cat obtained  by crushing
again each of the two components in Fig. \ref{tweezer}(b) between
ECs moving towards each other along the imaginary axis direction.

QZD is also obtained when the interrogation pulse has a
Rabi angle $\theta$ different from $2\pi$. The pulse performs then
a unitary kick acting on the atom/cavity system, which mixes
$|h,s\rangle$ with $|+,s\rangle$ and would create atom-field
entanglement if $C$ contained $s$ photons. This unitary admits an
invariant subspace, belonging to the eigenvalue +1 spanned by the
projection $|h\rangle\langle h|\otimes (P_{<s}+ P_{>s})$, the same
as for a $2\pi$ pulse. Starting from an atom in $|h\rangle$ and a
field in ${\cal H}_{<s}$ or ${\cal H}_{>s}$, we obtain a QZD
leaving the atom in $|h\rangle$ and the field in its initial
subspace. Under perfect QZD, the cavity \textsl{never} contains $s$
photons and the atom and field are \textsl{never} entangled by the
interrogation pulse. Obviously, QZD is not achieved if $\theta$ is
very small, each kick operation being too close to $\openone$. We have
checked numerically that, for $\theta\simeq 1$, we recover within
a good approximation all the results described above.

Numerical simulations can also take into account realistic
experimental imperfections. We have simulated a typical microwave
CQED experiment~\cite{Wigner08} with a cavity damping time
$T_c=0.13$~s and an atom-field coupling $\Omega/2\pi=50$~kHz. We
have taken into account the limited selectivity of the
interrogation pulse, which must be at the same time much shorter
than $T_c$ and much longer than $1/\Omega$. We have carefully
optimized the interrogation pulse parameters~\cite{Future} for a
tweezer operation leading, in 10 steps, from an
$|\alpha=-2\rangle+|\alpha=2\rangle$ cat to
$|\alpha=-3\rangle+|\alpha=3\rangle$. The total duration of the
simulated experiment is 3.4~ms and the final fidelity with respect
to the ideal cat is 76\%.

Quantum Zeno dynamics applied to a field oscillator leads to novel
methods for tailoring non-classical fields. Phase space tweezers
are promising tools for generating and manipulating arbitrary
superpositions of coherent states (i.e. arbitrary states
superpositions)~\cite{Future} and to study their evolution under
the effect of decoherence~\cite{Decoherence96}. Experimental
demonstrations of these effects are within reach of a state-of-the-art 
microwave CQED set-up. They could also be implemented in
circuit QED, a field in which high $\Omega T_c$ quality factors
are also realized~\cite{Schoelkopf10}. Manipulating at will the
state of a quantum oscillator in its phase space provides a new
insight into the physics of mesoscopic quantum superpositions and
the exploration of the quantum to classical boundary.

\begin{acknowledgments}
We acknowledge support by the EU and ERC (AQUTE and DECLIC projects) and by the ANR (QUSCO-INCA).
\end{acknowledgments}

\end{document}